# Spin Susceptibility and Specific Heat in the $d$-$p$ Model

Shigeru KOIKEGAMI* and Kosaku YAMADA[1]

*Institute for Solid State Physics, University of Tokyo, Kashiwa 277-8581*
[1]*Department of Physics, Kyoto University, Kyoto 606-8502*

We analyze the two-dimensional $d$-$p$ model, considering both antiferromagnetic spin fluctuation and $d_{x^2-y^2}$-wave superconducting fluctuation. We adopt the fluctuation-exchange approximation in order to derive both normal and anomalous vertices composed only of the renormalized $d$-electron Green function and the on-site repulsive interaction among the $d$-electrons. Using these vertices, we derive a $t$-matrix as a superconducting fluctuation propagator. Then, we obtain self-consistent solutions in which the system is close to antiferromagnetic instability. In our solutions, the superconducting fluctuation couples strongly with the quasiparticle state and this causes the anomalous behavior in the temperature dependences of spin susceptibility and specific heat.

The presence of a pseudogap in the normal state of underdoped high-$T_c$ cuprates (HTSCs) has been a significant issue in recent years. Judging from the various experiments on both the electronic states and the spin correlations, it seems natural that the pseudogap is due to the thermal superconducting fluctuation. When we study the realistic system, we need to treat both antiferromagnetic and superconducting fluctuation effects in a self-consistent manner. The approach proposed in ref. 1 is a natural extension of the fluctuation-exchange (FLEX) approximation, and it includes the superconducting fluctuation effect within the $t$-matrix approximation.

We apply this approach to a model with only repulsive interaction. Our model Hamiltonian is a two-dimensional $d$-$p$ model. It possesses only on-site Coulomb repulsion among $d$-electrons, $U$, as its interaction terms. Thus, our Hamiltonian is expressed as

$$H = \sum_{\bm{k}\sigma} \begin{pmatrix} d^{\dagger}_{\bm{k}\sigma} & p^{x\dagger}_{\bm{k}\sigma} & p^{y\dagger}_{\bm{k}\sigma} \end{pmatrix}$$
$$\times \begin{pmatrix} \varepsilon_d - \mu & \zeta^x_{\bm{k}} & \zeta^y_{\bm{k}} \\ -\zeta^x_{\bm{k}} & -\mu & \zeta^p_{\bm{k}} \\ -\zeta^y_{\bm{k}} & \zeta^p_{\bm{k}} & -\mu \end{pmatrix} \begin{pmatrix} d_{\bm{k}\sigma} \\ p^x_{\bm{k}\sigma} \\ p^y_{\bm{k}\sigma} \end{pmatrix}$$
$$+ \frac{U}{N} \sum_{\bm{k}\bm{k}'} \sum_{\bm{q}(\neq 0)} d^{\dagger}_{\bm{k}+\bm{q}\uparrow} d^{\dagger}_{\bm{k}'-\bm{q}\downarrow} d_{\bm{k}'\downarrow} d_{\bm{k}\uparrow} - \frac{U}{4} N \langle n_0^d \rangle^2 . \quad (1)$$

Here, $d_{\bm{k}\sigma}(d^{\dagger}_{\bm{k}\sigma})$ and $p^{x(y)}_{\bm{k}\sigma}(p^{x(y)\dagger}_{\bm{k}\sigma})$ are the annihilation (creation) operators for the $d$- and $p^{x(y)}$-electrons of the momentum $\bm{k}$ and spin $\sigma$, respectively. $\mu$ is the chemical potential. $\varepsilon_d$ is the site energy of the $d$-orbital measured from that of the $p$-orbital. We take the lattice constant of the square lattice formed on Cu sites as the unit of length; therefore, we can represent $\zeta^{x(y)}_{\bm{k}} = 2\,it_{dp} \sin\frac{k_{x(y)}}{2}$ and $\zeta^p_{\bm{k}} = -4t_{pp} \sin\frac{k_x}{2} \sin\frac{k_y}{2}$, where $t_{dp}$ and $t_{pp}$ are the transfer energies between a $d$-orbital and a neighboring $p^{x,y}$-orbital and between a $p^x$-orbital and a $p^y$-orbital, respectively. Hereafter, we take $t_{dp}$ as the unit of energy. The first term in eq. (1) is the unperturbed part. From this term we derive the unperturbed Green functions. The last two terms in eq. (1) are the interaction terms. The last one is the Hartree term; $N$ and $\langle n_0^d \rangle$ denote the number of Cu sites and the average number of $d$-electrons per site in the unperturbed state, respectively. When we consider the $d$-electron normal self-energy, we define $\varepsilon_d$ so that it includes the constant energy shift, $U\langle n_0^d \rangle/2$, and omit the Hartree term.

In our calculation, the $t$-matrix is calculated by the Bethe-Salpeter equation for the anomalous vertex. This vertex functions as the source of the attractive interaction in the $d_{x^2-y^2}$-wave Cooper pairing. Therefore, the $t$-matrix affects the $d$-electron self-energy as the superconducting fluctuation propagator. The anomalous vertex is derived by the FLEX approximation along with the normal vertex. The normal vertex directly contributes to the $d$-electron self-energy. Its contribution reflects the antiferromagnetic fluctuation effect on the electronic state. The derivation of these two types of vertices in the FLEX approximation and details of the numerical calculation are explained in ref. 2. In this paper, we investigate the temperature dependence of the spin susceptibility and that of the specific heat. Our calculation clearly shows that the doping dependences of these quantities are in good accordance with experimental results.

We execute our calculations for different values of the parameters, chemical potential, $\mu$, and temperature, $T$, with all other parameters fixed. Table I lists the fixed parameters. $W$ is the bandwidth of the highest band in the noninteracting state. Table II lists the final results at which the self-consistent convergence was reached. Here, $n_{d(p)}$ is the $d(p)$-hole number, and $\delta$ is defined by $\delta = n_d + n_p - 1$. Compared with the experimental values in Nuclear Magnetic Resonance (NMR) and Nuclear Quadrupole Resonance (NQR),[3] our results can be matched with $YBa_2Cu_3O_{7-x}$. When we investigate the temperature dependences of the physical quantities, we decrease $T$ with fixed $\mu$. At our lowest temperature, $T = 0.00696$, and $\delta$ decreases by 0.03 from the values at $T = 0.01392$ for both $\mu$s. However, we can consider the case for $\mu = 2.85$ as the *underdoped* case and the case for $\mu = 2.82$ as the *overdoped* case, because the difference in their $\delta$ is maintained as 0.04 throughout the temperature changes.

---

* E-mail address: skoike@issp.u-tokyo.ac.jp

In Fig. 1 we show the temperature dependences of $\{U\chi(q)\}_{max}$ and $\{\lambda(q)\}_{max}$. The antiferromagnetic or $d_{x^2-y^2}$-wave superconducting instability occurs when $\{U\chi(q)\}_{max} = 1$ or $\{\lambda(q)\}_{max} = 1$, respectively. In the underdoped case, i.e., for $\mu = 2.85$, our system becomes closer to having antiferromagnetic instability, that is, the antiferromagnetic fluctuation becomes stronger. In this case, the superconducting fluctuation mediated by the antiferromagnetic fluctuation couples more strongly with the one-particle states of $d$-electrons. Therefore, we can expect somewhat different temperature evolutions between these two cases.

| $\varepsilon_d$ | $t_{pp}$ | $W$ | $U$ |
|---|---|---|---|
| 1.280 | 0.350 | 2.889 | 2.900 |

Table I. Set of parameters for the numerical calculation.

| $(\mu)$ | $n_d$ | $n_p$ | $\delta$ | $n_d/n_p$ |
|---|---|---|---|---|
| (2.82) | 0.713 | 0.493 | 0.210 | 1.45 |
| (2.85) | 0.706 | 0.467 | 0.173 | 1.51 |

Table II. Results for the numerical calculation at $T = 0.01392$.

Here, we study the temperature dependences of the spin correlations. First, in Fig. 2 we show the imaginary part of $\chi^{RPA}(\boldsymbol{q},\omega) \equiv \frac{\chi(\boldsymbol{q},\omega)}{1-U\chi(\boldsymbol{q},\omega)}$ on the line in momentum space, $\boldsymbol{q} = (\pi, q_y)$. Here, $\chi(\boldsymbol{q},\omega)$ is the bare spin susceptibility of $d$-electrons. Comparing (a) with (b), it is evident that $Im \chi^{RPA}(\pi, q_y, \omega)$ has the larger incommensurate peaks at $\omega \simeq 0.010$ for the lower temperature. $Im \chi^{RPA}(\boldsymbol{q},\omega)$ determines the strength of the attractive interaction for the $d_{x^2-y^2}$-wave superconductivity. Therefore, this behavior of $Im \chi^{RPA}(\boldsymbol{q},\omega)$ is consistent with the finding that $\{\lambda(q)\}_{max}$ is more enhanced at the lower temperature, as shown in Fig. 1.

The superconducting fluctuation has a smaller characteristic energy than the antiferromagnetic one. Therefore, its effect on the spin dynamics appears in the low energy limit. In order to confirm this nature, we calculate the spin-lattice relaxation rate in NMR, $(T_1T)^{-1}$. If we neglect the momentum dependence of the hyperfine coupling constant and the proportional coefficient,

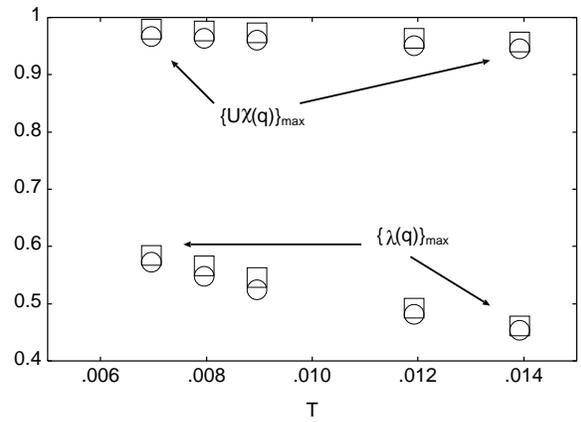

Fig. 1. $\{U\chi(q)\}_{max}$ and $\{\lambda(q)\}_{max}$. Open squares are for $\mu = 2.85$, and open circles are for $\mu = 2.82$.

$(T_1T)^{-1}$ relates to $Im \chi^{RPA}(\boldsymbol{q},\omega)$ as

$$(T_1T)^{-1} = \lim_{\omega \to 0} \sum_{\boldsymbol{q}} \frac{Im \chi^{RPA}(\boldsymbol{q},\omega)}{\omega}. \qquad (2)$$

Since $Im \chi^{RPA}(\boldsymbol{q},\omega)$ has a prominent value only around the incommensurate peaks, $(T_1T)^{-1}$ reflects the low energy state around these peaks. Furthermore, from the wave vectors $(\pi, \pi \pm \delta_{inc})$ corresponding to these peaks it is evident that the quasiparticle states around $(\pm\pi, 0)$ are connected with the ones around $(0, \mp\pi)$. These points are most subject to the $d_{x^2-y^2}$-wave superconducting fluctuation effect. Consequently, the superconducting fluctuation effect appears most significantly in the behaviors of $(T_1T)^{-1}$. In Fig. 3(a) we present the temperature dependence of $(T_1T)^{-1}$. $(T_1T)^{-1}$ has the maxima in the normal phase, as observed in the NMR experiments for underdoped HTSCs.[4-7] The superconducting fluctuation competes with the antiferromagnetic one and causes the maximum of $(T_1T)^{-1}$ relative to temperature. Comparing the two cases, we can find that the more underdoped case has a higher maximum temperature, $T^* = 0.008 \simeq 80K$, than the other, $T^* \simeq 75K$. This result is due to the different strength of the attractive interaction, i.e., the antiferromagnetic fluctuation. As the antiferromagnetic fluctuation grows stronger, the superconducting fluctuation more firmly couples $d$-electrons more strongly. As a result, the change of the magnetic excitation appears at higher temperatures even if both systems are far from the Cooper instability.

The uniform static spin susceptibility, $\chi^{RPA}(0)$, is presented in Fig. 3(b). For both cases, $\chi^{RPA}(0)$ decreases with decreasing temperature. Comparing the two cases, we can see that the more underdoped case has less susceptibility. In the NMR experiments, these results are observed as the thermal contraction of the Knight shift, which includes the part which is proportional to the uniform static spin susceptibility.[7]

Next, we investigate other information related to the quasiparticle state around the Fermi surface. The total density of state at the Fermi level $\rho(0)$, and the electronic specific heat coefficients $\gamma = C_{el}/T$ yield the integrated quantities on the Fermi surface. For the $d$-$p$ model, Kohno and Yamada analyzed $\gamma$ on the basis of Fermi liquid theory.[8] We adopt their definition, and, except for the proportional constant, $\gamma$ is given as

$$\gamma = \int d\boldsymbol{k} \left[ \rho_{\boldsymbol{k}}^{d}(0) \left( 1 - \left.\frac{\partial \Sigma(\boldsymbol{k},\omega)}{\partial \omega}\right|_{\omega=0} \right) + \rho_{\boldsymbol{k}}^{p^x}(0) + \rho_{\boldsymbol{k}}^{p^y}(0) \right]. \quad (3)$$

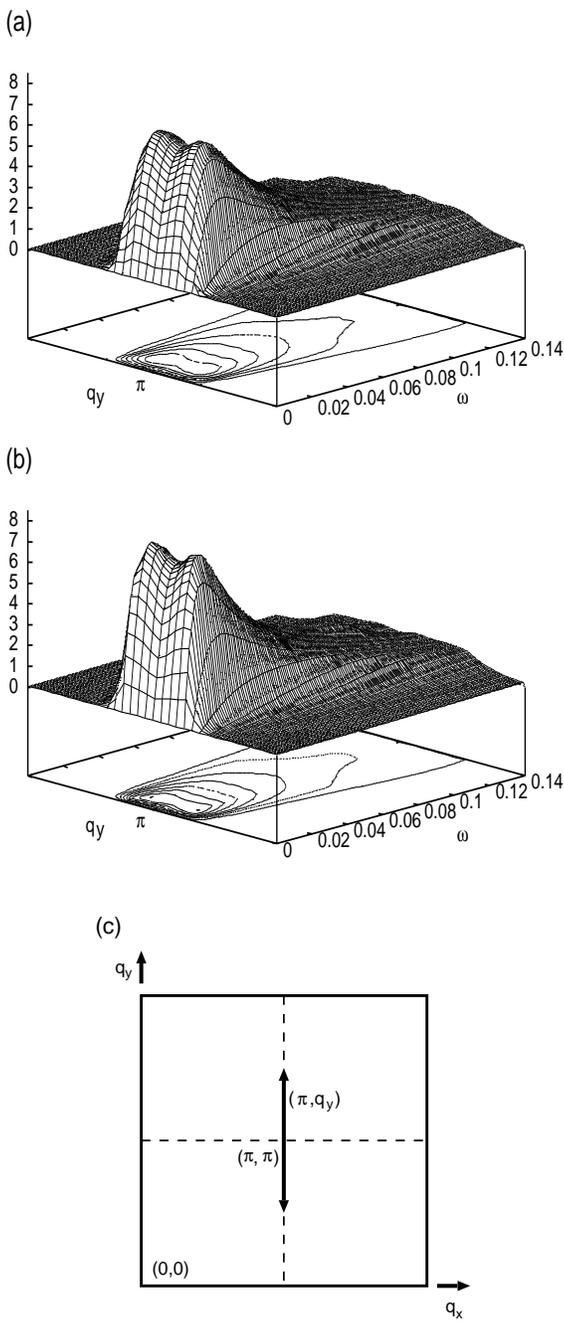

Fig. 2. Im $\chi^{\rm RPA}(\pi, q_y, \omega)$ for the underdoped case, i.e., $\mu = 2.85$, (a) at $T = 0.00796$ and (b) at $T = 0.00696$. (c) shows the line in the two-dimensional momentum space, where we scan Im $\chi^{\rm RPA}(\pi, q_y, \omega)$ in (a) and (b).

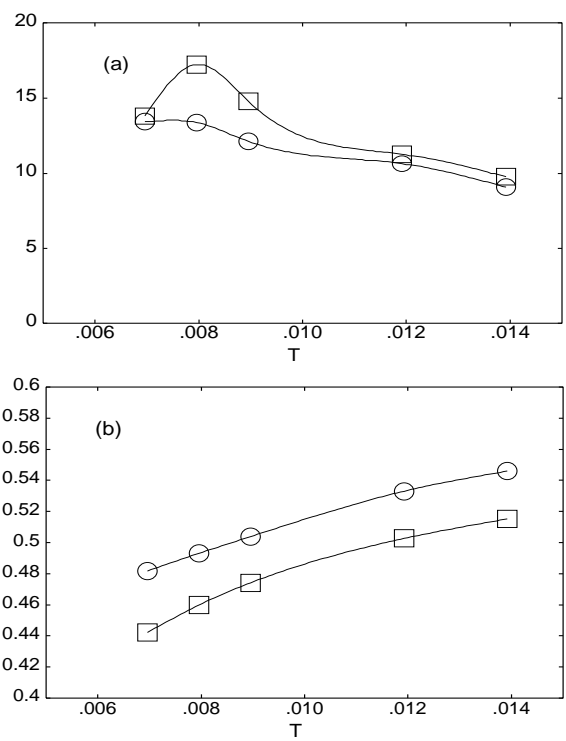

Fig. 3. Temperature dependences of $(T_1T)^{-1}$ (a) and $\chi^{\rm RPA}(0)$ (b). Open squares represent $\mu = 2.85$, and open circles represent $\mu = 2.82$. Solid lines are visual guides.

Here, $\rho_{\boldsymbol{k}}^{d(p^\alpha)}(0) = -\frac{1}{\pi} \operatorname{Im} G_{d(p^\alpha)}(\boldsymbol{k},0)$ ($\alpha = x, y$). $\rho(0)$ is defined as

$$\rho(0) = -\frac{1}{\pi} \operatorname{Im} \int d\boldsymbol{k} \operatorname{Tr} \hat{G}(\boldsymbol{k},0). \quad (4)$$

In Figs. 4(a) and (b) we show the temperature dependences of $\rho(0)$ and $\gamma$, respectively. Since we fixed the chemical potential $\mu$, the density of state at the Fermi level should tend to increase with decreasing temperature. However, for the underdoped case, i.e., $\mu = 2.85$, we can recognize that both $\rho(0)$ and $\gamma$ decrease with the temperature for $T \leq 0.0120$. In the underdoped case, stronger antiferromagnetic fluctuation suppresses the density of state around the Fermi surface, compared with the overdoped case, i.e., $\mu = 2.82$. Since the antiferromagnetic fluctuation has the characteristic energy $\omega \simeq 0.010$, the declines of both $\rho(0)$ and $\gamma$ start at a high temperature. Comparing Fig. 4(a) with 4(b), we see that below $T = 0.008$, $\gamma$ declines more rapidly than $\rho(0)$. This is mainly due to the decrease of $\gamma_{\boldsymbol{k}} \equiv 1 - \left.\frac{\partial \Sigma(\boldsymbol{k},\omega)}{\partial \omega}\right|_{\omega=0}$. As well as in $(T_1T)^{-1}$, the superconducting fluctuation effect competes with the antiferromagnetic one in $\gamma_{\boldsymbol{k}}$. These temperature dependences of $\gamma$ for underdoped cuprates have been observed in some well-done experiments.[9,10] Thus, our calculation explains the reduced specific heat in the underdoped case.

We have carried out an almost fully self-consistent calculation on the basis of the two-dimensional $d$-$p$ model, which possesses only repulsive interaction among $d$-electrons. In the treatment that considers only the antiferromagnetic fluctuation, the system becomes so close to having antiferromagnetic instability that it was difficult to obtain the self-consistent solutions in the underdoped states. If we consider the superconducting fluctuation as well, the antiferromagnetic correlation decreases due to the suppression of the density of states around the Fermi level. Thus, we can obtain the self-consistent solutions over a wide temperature range in the underdoped states.

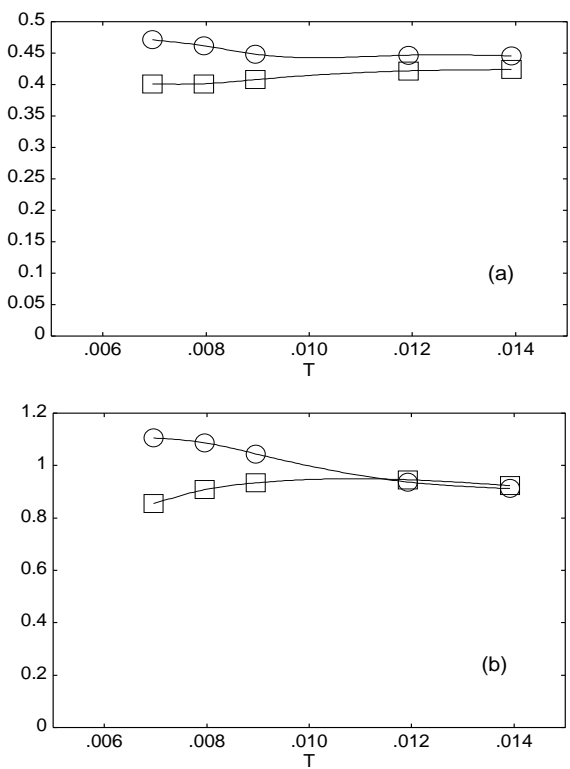

Fig. 4. Temperature dependences of $\rho(0)$ (a) and $\gamma$ (b). Open squares represent $\mu = 2.85$, and open circles represent $\mu = 2.82$. Solid lines are visual guides.

Our solutions have successfully explained various experimental results of the normal phase of HTSCs, e.g., the anomalous temperature dependence of $(T_1T)^{-1}$ in the NMR experiment; in the temperature evolution of $(T_1T)^{-1}$, the superconducting fluctuation plays a role in decreasing it. Each of the two types of fluctuations competes with the other in the low energy region. This competition brings $(T_1T)^{-1}$ to the maximum in its temperature evolution. The electronic specific heat coefficient $\gamma$ is also affected by the superconducting fluctuation at low temperature as well as the antiferromagnetic fluctuation effect beginning at the higher temperature. As indicated by the different behaviors of these quantities between the filling states, the effect of the superconducting fluctuation is determined by the strength of the antiferromagnetic fluctuation. When the antiferromagnetic fluctuation is weak, the superconducting fluctuation cannot function strongly.

Both spin and charge dynamics of quasiparticles can be affected by the antiferromagnetic and superconducting fluctuations. However, the characteristic energy ranges for the two types of fluctuation are somewhat different. The superconducting fluctuation effect is relevant to the lower energy part, but irrelevant near the characteristic spin fluctuation energy. The antiferromagnetic fluctuation remains strong even if the superconducting fluctuation effect occurs. On the other hand, the antiferromagnetic fluctuation is the source of the attractive interaction for $d_{x^2-y^2}$-wave Cooper pairs. Therefore, even at higher temperatures, the superconducting fluctuation can occur because its effect is irrelevant to the antiferromagnetic fluctuation. Moreover, the low-dimensionality characteristic of HTSCs increases the fluctuation effects on the quasiparticle state. Thus, in the underdoped state the enhanced antiferromagnetic fluctuation firmly binds the superconducting fluctuation to quasiparticles.

In conclusion, we can explain the anomalous temperature dependences of $(T_1T)^{-1}$ in NMR and the electronic specific heat coefficient for the underdoped cuprates by considering the superconducting fluctuation effect. These anomalous behaviors originate from the competition of the antiferromagnetic fluctuation with the superconducting fluctuation. This competition should lead to superconductivity with the antiferromagnetic instability suppressed at lower temperatures. We can expect a more marked change in the spin and charge dynamics of quasiparticles.

The authors are grateful to Professor M. Ido, Dr. N. Momono, Dr. K. Ishida, Dr. S. Nakamura, Dr. T. Takimoto, Dr. A. Kobayashi, Dr. R. Arita, Mr. S. Onoda, Mr. Y. Yanase and Mr. T. Jujo for fruitful discussions. The computation in this work was performed using a Fujitsu VPP500 at the Supercomputer Center, Institute for Solid State Physics, University of Tokyo. The computation was also executed using a Pilot-3 of the Research Center for Computational Physics, Tsukuba University. One of the authors (S. K.) thanks Mr. M. Matsubara for useful advice regarding the computation on the Pilot-3. S. K. was supported by Research Fellowships of the Japan Society for the Promotion of Science for Young Scientists.